\documentclass[11pt]{article}

\usepackage{myArticle}
\usepackage{lscape,multicol,natbib}
\usepackage{booktabs}
\hypersetup{breaklinks=true}

\usetikzlibrary{arrows,backgrounds,calc,decorations.pathreplacing,decorations.pathmorphing,positioning,shapes.misc}
\tikzset{snake it/.style={decorate, decoration={snake, amplitude=0.3mm, segment length=2mm}}}

\title{Towards a Sustainable Power Grid: Stochastic Hierarchical Planning for High Renewable Integration}
\author[1]{Semih~Atakan\thanks{atakan@usc.edu}}
\author[2]{Harsha Gangammanavar\thanks{harsha@smu.edu}}
\author[1]{Suvrajeet Sen\thanks{s.sen@usc.edu}}
\affil[1]{Department of Industrial and Systems Engineering, University of Southern California, Los Angeles, CA}
\affil[2]{Department of Operations Research and Engineering Management, Southern Methodist University, Dallas TX}
\date{\today ~(First version: 09/2019)\footnote{The first version was entitled ``Operations planning experiments for power systems with high renewable resources.''}} 

\newcommand{\sa}[1]{#1}
\newif\ifpreprint
\preprinttrue

\begin{document}\thispagestyle{empty}
\maketitle

\begin{abstract}
Driven by ambitious renewable portfolio standards, large scale inclusion of variable energy resources (such as wind and solar) are expected to introduce unprecedented levels of uncertainty into power system operations. The current practice of operations planning with deterministic optimization models may be ill-suited for a future with abundant uncertainty. To overcome the potential reliability and economic challenges, we present a stochastic hierarchical planning (SHP) framework. This framework captures operations at day-ahead, short-term and hour-ahead timescales, along with the interactions between the stochastic processes and decisions. In contrast to earlier studies where stochastic optimization of individual problems (e.g., unit commitment, economic dispatch) have been studied, this paper studies an integrated framework of \emph{planning under uncertainty}, where stochastic optimization models are stitched together in a hierarchical setting which parallels the deterministic hierarchical planning approach that is widely adopted in the power industry. Our experiments, based on the NREL-118 dataset, reveal that under high renewable integration,  significant operational improvements can be expected by transitioning to the SHP paradigm.  In particular, the computational results show that significant improvements can be achieved in several metrics, including system reliability, environmental sustainability, and system economics, solely by making a strategic choice to adopt the new SHP paradigm. 
\end{abstract}

\section{Introduction} \label{sect:intro}
Lawmakers throughout the U.S.\ have mandated that renewable resources meet a significant percentage of the electricity supply. Each state has set its own goals, with California being the most aggressive, requiring 50\% renewables by 2026, 60\% by 2030, and 100\% by 2045 \citep[see][]{Proctor2018}. European countries have also set ambitious targets both at the level of individual nations, as well as the European Union as a whole \citep{Irena2018}. State and local authorities (e.g., Independent System Operators (ISOs) in the U.S. and Transmission System Operators (TSOs) in Europe) have commissioned studies to assess operational considerations such as system reliability, market design, incorporation of storage technologies, and other avenues. A recent simulation study \citep{Olson2015}, commissioned by California ISO (CAISO), suggests that for renewable-integration levels beyond 33\%, one can expect a fair amount of over-generation and renewable curtailment during the daytime, and perhaps, load-shedding around sundown. Higher levels of renewable integration exacerbate these issues. 

\sa{Power systems operations planning processes are mainly based on a hierarchical strategy. This strategy leverages layers of optimization models solved in sequence using a rolling horizon approach, to improve the reliability and economics of power systems operations. Typically, at the top layers of this hierarchy, long-horizon and low-resolution models are executed to determine commitment plans for generators with slower-ramping capabilities (e.g., nuclear). These plans are then passed to downstream layers of the hierarchy, where the decision space is gradually refined with higher-resolution models. The higher-resolution models use more accurate forecasts for demand and renewable generation to determine the commitment plans of most generators. Finally, a few minutes before dispatch, the economic dispatch model is solved to refine the generation plan for the committed generators. Electricity supply in this setup can be thought of as an aggregation of several products supplied to the consumer via a ``middle-man'', the ISO/TSO. These entities use a portfolio of power contracts that obligates a producer to provide electricity at a cleared price during the corresponding time window. The market time windows are aligned with the operations in a similar manner. For instance, a low-resolution market in the day-ahead is followed by a higher resolution real-time market closer to the time of actual dispatch. The operational elements (generator commitments, generation amounts, etc.) and the market elements (electricity prices) are outputs of the rolling-horizon optimization models.

Today's power system planning process use \emph{deterministic} optimization models in all layers of the hierarchy. Such a process can be termed as deterministic hierarchical planning (DHP). The adoption of DHP for planning power systems operations predates the introduction of vast amounts of renewable energy into the grid. The significant levels of volatility of renewable resources pose one of the major challenges to maintaining reliable supply under DHP. Indeed, a recent study points to a significant increase in costs to maintain a fully renewable power system while meeting today's level of system adequacy \citep{Zappa2019}.}

\begin{figure}
\centering
\includegraphics[width=0.85\columnwidth]{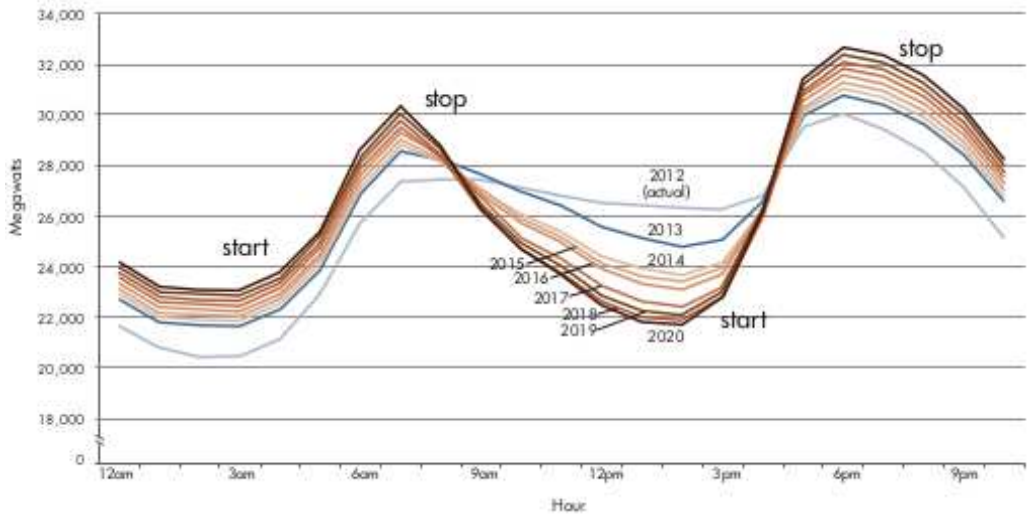}
\caption{CAISO's duck chart, predicting four emerging ramping patterns with increased renewable integration \cite{caiso2016}.} 
\label{ch:td:fig:duckchart}
\end{figure}

A popular illustration of the issues with renewable integration is captured by the so-called ``duck-chart'' of CAISO (see Fig.\ \ref{ch:td:fig:duckchart}). This figure depicts the daily net-load (total electricity load minus generation from ``must-run'' units) across successive years with increasing levels of solar added to the generation mix. A surplus of solar energy during the daytime leads to a dip in the net load, followed by a significant upward ramp around the sundown. In a grid with limited storage capabilities, excess supply during daytime poses significant challenges as utilities will be required to procure sufficient ramp-up capabilities to meet the electric load during evening hours. The absence of substantial ramping capabilities can push the loss-of-load probability to unacceptable levels, and may even cause load-shedding in certain areas, jeopardizing system reliability and performance. On the other hand, over-generation during daytime could lead to negative prices in the market, resulting in, for instance, large shipments of energy to neighboring states (e.g., from California to Arizona) while paying these states to accept the surplus \citep[see][]{Penn2017}. \sa{Additionally, it is important to recognize that Fig.\ \ref{ch:td:fig:duckchart} is based on point forecasts and overlooks the uncertainty in demand and renewable generation. During actual operations, the trajectory of the net-load is unknown to the operators and can significantly deviate from the point forecast. Therefore, plans based on deterministic models that use point forecasts as input will result in immense reliability challenges. }

To overcome the challenges of renewable integration and ``tame the duck'', so to speak, a recent U.S.\ Department of Energy (DOE) report \citep{Bergman2016} has distilled a myriad of operational guidelines (for maintaining reliability) into four specific rules: (1) Power generation and transmission capacity must be sufficient to meet the peak demand for electricity; (2) Power systems must have adequate flexibility to address variability and uncertainty in demand and generation resources; (3) Power systems must be able to maintain a steady frequency; (4) Power systems must be able to maintain a steady voltage at various points on the grid. The focus of these rules is on expected changes over the next several years due to the inclusion of a new generation of resources, especially variable energy resources (VER). The first two rules address operations \textit{planning}, whereas the last two rules pertain to operations \textit{control}.
It is fair to suggest that technological solutions for operations control, which permit high renewable integration, are well underway. 


As for operations planning, we posit that it is necessary to reassess the structure of the DHP paradigm to accommodate the volatility associated with renewable energy that affects every layer of this hierarchy. Such an assessment involves studying alternative setups that allow combinations of contemporary (deterministic) planning models and novel (stochastic) planning models.  Given this focus, the main contributions of this paper are as follows.
\begin{enumerate}
    \item \textit{Overarching Goal: Comparing Stochastic and Deterministic Hierarchical Planning.} We study the implications of transitioning from DHP towards \emph{stochastic hierarchical planning} (SHP) in a system with a centralized planner. We investigate whether it is possible to fully realize the positive impacts of VER integration (i.e., reduced greenhouse gas (GHG) emissions) without deterioration in system reliability and significant increases in production and distribution costs. The novelty of our work is in an integrated assessment of combinations of models that are arranged in a hierarchy, and aware of system volatility due to VERs. 
    
    \item \sa{\textit{A Decision Evolution Architecture.} The decision processes used in this paper are tightly connected with the NREL-118 dataset \citep{NREL118}, which not only provides the grid data, but also the renewable energy environment for our experiments. In particular, the evolution of decisions and uncertainty allowed by this dataset lets us investigate the goal set in item 1. Our architecture allows us to draw conclusions regarding costs, GHG emissions, load shedding, etc., based on the specific decision-modeling choices.}
    
    
    \item \textit{Experimental Study.} We perform a comprehensive computational study of the SHP under different models (deterministic v.\ stochastic) at individual layers of the hierarchy, varying renewable integration levels, and system reliability requirements. Our computational study demonstrates the potential of SHP when compared to the current DHP strategy and acts as a precursor to an ISO-scale study.
\end{enumerate}

\sa{It is important to emphasize that this study focuses on the NREL-118 dataset, and while similar studies can be undertaken using other datasets, the conclusions of this paper pertain only to those power systems which have similar VER levels and weather patterns. Our thesis is that the choice of the decision hierarchy of a system should be made in a manner that reflects the renewable integration levels because systems with high VERs may be better served using SHP, while those with lower VERs may continue to adopt the DHP setting. To give the reader a brief preview of our results, it turns out that the current DHP framework may not be as effective for decision-making under uncertainty. We present strong evidence that for NREL-118, SHP provides superior metrics for operational costs, lost-load, as well as GHG emissions.} 

Studies of using stochastic optimization for power system operations can be traced back at least forty years to \cite{MSSoyster1982}. While there are many studies of stochastic optimization within any one layer of the hierarchical system, no other study pits the standard hierarchy of deterministic models against a hierarchy of stochastic ones in a system operated by a centralized planner. It is such a comparison that provides a preview of the potential advantages and disadvantages of alternative hierarchies (DHP v.\ SHP) on electricity production planning. Thus, this study examines whether a system-wide overhaul that introduces stochastic optimization and coordination among one or more layers of the hierarchy can mitigate difficulties associated with high integration of renewable energy into the grid. 

The remainder of this paper is organized as follows. We begin with an overview of today's power system operations in \S\ref{sect:overview} and discuss relevant literature in \S\ref{sec:literature}. In \S\ref{sec:decisionProcess} we present a detailed description of the SHP framework, including optimization models and solution algorithms used to simulate system performance. In \S\ref{sec:experiments} we present our experimental results. Finally, we conclude with a brief discussion of how one might proceed to the next phase of tests, which would require ISO-size networks and technologies in \S\ref{sec:conclude}. The overall structure of each model used in the hierarchy is discussed in Appendix A, and details on solar and wind inputs are given in Appendix B. In the interest of enhancing reproducibility of our study, the code and data associated with our experiments are available at \cite{github}.

\section{An Overview of Power Systems Operations Planning} \label{sect:overview}

Electric power systems are very large-scale networks interconnecting many sources of electric power to points of consumption. The entire network is arranged at several voltage levels, converted from one to the other by step-up or step-down transformers. Minimizing total cost while ensuring reliable power delivery to customers is the overarching goal of the system operations. Given the continuous and very large-scale nature of operations, the implementation of this objective is complex when viewed as a single decision-making problem. Therefore, system operators use a reformulation involving a hierarchy of optimization models defined over overlapping horizons with different time resolutions for decisions and constraints.

Our hierarchical planning paradigms assume coordination via a central planning authority (e.g., ISO or TSO). Most ISOs/TSOs currently implement some form of a DHP framework, which divide the daily planning activities into three principal layers: a) \emph{day-ahead unit commitment} based on a daily forecast of load and generation limits leading to a production and transmission plan, b) \emph{short-term unit commitment} over a shorter planning window (typically three to four hours), producing some commitment decisions and updated transmission plans, and c) \emph{hour-ahead economic dispatch} where the production and transmission plans are finalized and necessary reserve capacities are committed. There are some variations of this multi-layer hierarchy, such as updating the dispatching plan in 15-minute intervals to accommodate high levels of VER. However, all layers in this hierarchical setup adopt some form of deterministic optimization models, and as such, all forecasts used in these models are \emph{point forecasts}. In what follows, we give further details on the three principal layers mentioned above, and provide notes on how our setup (described in greater depth later) incorporates these layers. 

\subsubsection*{Day-ahead (DA) Operations} This phase begins by estimating demand and renewable-supplies, as well as collecting generation and demand bids via electricity markets. This information is used in simultaneous co-optimization for the next operating day, using security-constrained unit commitment (UC) and security-constrained economic dispatch (ED) models. In most settings (including ours), these models are formulated over a $24$-hour horizon, with decisions and constraints specified at an hourly resolution. The UC model commits and schedules resources for regulation. The amount of resources (mainly spinning operating reserves) scheduled in DA are based on estimates generated using historical data, as well as ISO-specific practices. DA planning also involves committing resources for reliability assessment and emergency operations. For simplicity, we do not consider these in our setup. The UC optimization model involves continuous as well as binary decision variables, resulting in a mixed-integer program (MIP).

The UC decisions are used to instantiate the DA security-constrained ED model. This model determines generation, regulating and spinning reserve amounts for all committed resources, as well as ex-ante DA prices. While the ED model in the DA phase is often solved separately for each hour of the day, we formulate the ED model as a single optimization model defined over an entire day at an hourly resolution. In our models, we do not allow for generator self-scheduling and do not consider system operations under contingency/emergency.

\subsubsection*{Short-term (ST) Operations} 
Some of the advanced ISOs use additional instruments that commit fast-start resources to ensure schedules meet all reliability requirements. The associated models are solved independently against DA transactions and generation bids. At certain ISOs (e.g., NYISO), these operations are considered to be part of the real-time (RT) markets and are referred to as the RT-UC. Following the terminology at CAISO, we will refer to these operations as Short-Term UC (ST-UC). These models are formulated at finer resolution (e.g., $15$ mins), allow adaptive (de)commitment decisions, and solved over a horizon of few hours (e.g., $4.5$ hrs at CAISO and $2.5$ hrs in NYISO). In our setup, we define these models at a resolution of $15$ minutes and a horizon of $4$ hours, solved every $3$ hours.

\subsubsection*{Real-time (RT) Operations} There is always some RT deviation of actual generation and load from what was scheduled during DA planning. One of the key functions of the ISO is to perform real-time balancing of loads and generation. RT balance is maintained through a combination of spinning and ancillary services along with the units providing regulation reserves, which are managed by automatic generation control (AGC). The non-AGC units are dispatched every few minutes (usually $5$-$15$ mins), while the regulation units are used only to respond instantaneously to system imbalances. The planning of dispatch amounts for RT operations are done using ED models with 5 to 15 minute intervals. Our setup will consider ED models at a 15-minute resolution with 75-minute horizon, but set aside tasks associated with downstream models (managing actual dispatch and transmission control) as being outside the scope of this paper.

\sa{
\section{Literature Review} \label{sec:literature}
}
Even with the decomposition of operations planning into DA, ST and RT phases, generation scheduling and dispatch problems are truly stochastic optimization problems, therefore they are computationally very challenging. Under a DHP setting, system operators approximate the random variables with their point forecasts, converting these stochastic problems into large-scale deterministic (mixed-integer) problems, which can be handed off to commercial solvers. Note that electricity production and distribution are governed by nonlinear equations which can lead to non-convex problems. For the sake of scalability, we focus on linear approximations of all such non-linear model components (similar to the literature cited in this section).

Both deterministic UC and ED problems are fundamental to power system operations planning, and both have been studied extensively in the literature. Comprehensive treatments of state-of-the-art UC formulations can be found in \cite{Knueven2019}, where the latter also includes a detailed computational study. The surveys \cite{Frank2012a, chowdhury1990review} provide an overview of deterministic ED models.

Typical UC formulations studied in the literature \citep[such as][]{Ostrowski2012, Atakan2018uc, Knueven2019} do not include a transmission network, and model dispatch at an aggregate level. 
In the interest of greater accuracy with transmission constraints, we adopt the dispatch model studied in \cite{Gangammanavar2015}. This formulation involves a linear objective function that captures production costs and over-generation/load-shedding penalties, as well as constraints corresponding to generation capacities, ramping, flow balance, linearized power flow (DC approximation), operating reserve utilization, bounds on bus angles, and line capacities.

The traditional approach for addressing uncertainty has been, in essence, over-producing, which creates a reserve of electricity that is used as a buffer against unexpected volatility. Some ISOs (e.g., New England ISO) have already recognized the shortcomings of deterministic planning within the context of renewable integration, and recommend certain deterministic policies \citep{Zhao2015}.
Over the last decades, developments in stochastic modeling and optimization have allowed uncertainty to be explicitly handled within the decision process. In particular, stochastic programming (SP) has played a prominent role to enable decision-making under uncertainty in real-scale problems across many application domains including power systems \citep{Wallace2005}. Two-stage stochastic programs (2-SPs), including models with discrete first-stage variables, have gained acceptance among both the power systems research community as well as practitioners.

Stochastic UC models have been studied to better accommodate wind uncertainty  \citep{Ruiz2009, Papavasiliou2011}; or other sources of stochasticity such as generator/transmission failures and demand uncertainty \citep{Zheng2013, Cheung2015}.
For a review paper on stochastic UC, we refer the reader to \cite{Zheng2015}. To account for uncertainty closer to the time of dispatch, stochastic ED models have been investigated using stochastic programming  \citep[e.g.,][]{Gangammanavar2015, gu2016stochastic} or robust optimization approaches \citep[e.g.,][]{lorca2014adaptive}.


In addition to deterministic v.\ stochastic comparisons of individual models, longer term impact of stochastic models are assessed in \cite{Tuohy2009, Sturt2012}. In contrast to earlier citations, these studies capture the temporal dynamics of system operations by solving sequences of stochastic planning models over a rolling time horizon \citep[i.e., model predictive control as in][]{Grossman2003}. Our setup is similar in this respect, as it also involves (multiple types of) models solved over a rolling time horizon. Under (supply) uncertainty, studies were conducted to bridge the gap between DA planning and RT execution \citep[e.g.,][]{SurenderReddy2016, Choi2018}. Both of these studies leverage deterministic linear optimization models and their focus is on coordinating DA and RT operations as opposed to assessing the impact of VER integration. 

In contrast, \cite{Ilic2007} calls for considering a hierarchical framework involving \emph{stochastic} dynamic problems evolving at different timescales.  While their paper provides high-level  (big-picture) assessment, it does not provide a quantitative assessment. On the other hand, \cite{Schulze2016} investigate what advantages might accrue from using stochastic optimization in stochastic DA and intra-day (ST) UC problems over a multi-year time horizon. However, it is important to highlight some of the differences between their setup, and ours.  First, their  study does not consider ED as part of the planning process, and moreover, their models and data may be described as a somewhat lower resolution representation which includes a much smaller set of scenarios in their stochastic programs. The dataset used in our study \citep{NREL118} was specifically introduced by NREL for the purpose of investigating systems with high VERs. Due to the level of detail provided in this dataset, as well as our focus on methods which allow a very large collection of scenarios, our conclusions happen to be different from that reported by \cite{Schulze2016}, whose experiments observed only modest improvements resulting from stochastic models with a very coarse description of scenarios and time resolutions.

It is worthwhile to note that prices in power systems are settled through hierarchically arranged markets which involve generator companies, consumers (i.e., loads) as well as transmission owners as their participants. Each participant (e.g., a generator company) bids their price and capacity, and the ISO/TSO is given the responsibility to clear the market. We emphasize that our study focuses on operational models which are instantiated by such centralized planners (e.g., ISO/TSO) using bids submitted via the market. Moreover, our study inherently assumes competitive markets, risk-neutral market players, and cost-optimal bid prices for all market participants. We refer the reader to \cite{conejo2010decision} for a detailed treatment of electricity markets under uncertainty; and \cite{Ralph2015} for treatment of risk-aware market participants. 

In the context of markets, there have been recent attempts to capture multiple layers at different timescales. \cite{Dorostkar2019} consider the IEEE-33 instance and adopt a two-layer hierarchy with DA and RT layers. 
\cite{Aasgard2019} provide a conceptual study where they argue the benefits of a multi-market (hierarchical) bidding process (mainly for hydro-generators). \cite{Ottesen2018} consider an optimal bidding problem 
over a three-layer energy market, where stochastic programs under price uncertainty are adopted at each layer. Finally, \cite{Morales2014} consider a two-layer energy market, analyze shortcomings of conventional and stochastic market clearing models and provide a bi-level programming solution.

Despite the significant attention paid to address the challenges of integrating large-scale renewable resources into power systems, the scope of the models considered is limited to a single layer in the hierarchy. The only exceptions are the studies noted above that mostly focus on electricity markets. Even when 2-SP models are used, the evaluation (using Monte Carlo simulations or against historical data in a rolling horizon setting) is limited to the first-stage decisions. In contrast, our setup allows us to evaluate decisions at \textit{all layers} considered in the hierarchy. In this regard, this setup closely mimics the actual operations of a power system.  We reiterate that this is the whole point behind using NREL-118 for our experimental design.


\section{Stochastic Hierarchical Planning} \label{sec:decisionProcess}
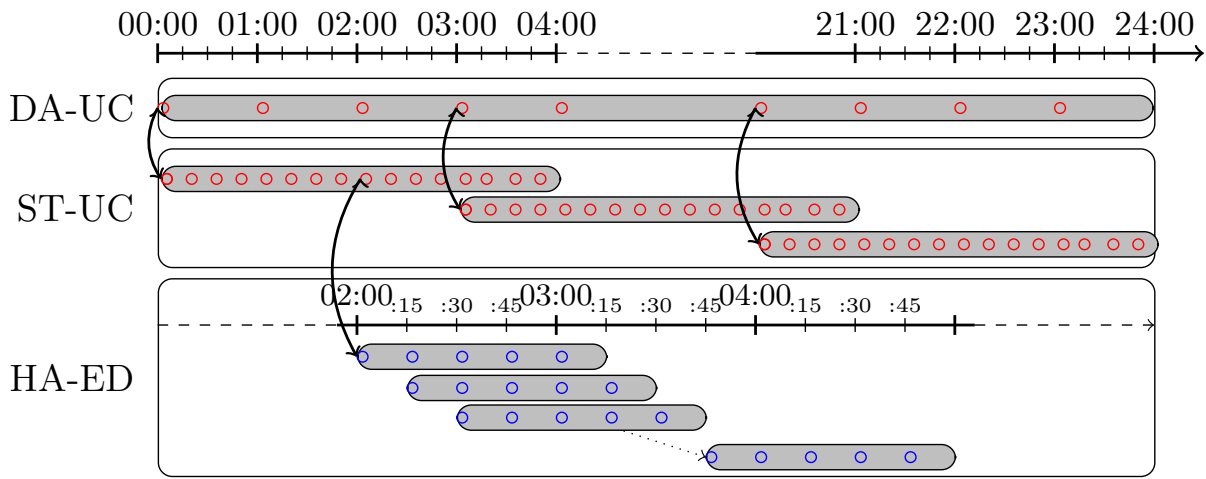
\begin{figure*}[ht]
		\begin{tikzpicture}[]
		\tikzset{timeTick/.style={font=\small},
			probType/.style={rectangle,draw,rounded corners, align=center, anchor=north west},
			textbox/.style={font=\scriptsize}}

		\pgfmathsetmacro{\y}{0}
		\draw[thick] (0,\y) -- (4,\y); \draw[dashed] (4,\y) -- (8,\y); \draw[thick,->] (6,\y) -- (10.5,\y);		
		\foreach \x in {0,1,2,3,4,7,8,9,10}{
			\draw[thick] (\x,\y-0.1) -- (\x,\y+0.1);	
		}
		\foreach \x in {0.25,0.5,0.75,1.25,1.5,1.75,2.25,2.5,2.75,3.25,3.5,3.75,7.25,7.5,7.75,8.25,8.5,8.75,9.25,9.5,9.75}{
			\draw (\x,\y-0.07) -- (\x,\y+0.07);
		}
		\node[timeTick] at (0,\y+0.3) (startTime) {00:00}; 
		\node[timeTick] at (1,\y+0.3) {01:00}; 
		\node[timeTick] at (2,\y+0.3) {02:00}; 
		\node[timeTick] at (3,\y+0.3) {03:00}; 
		\node[timeTick] at (4,\y+0.3) {04:00}; 
		\node[timeTick] at (7,\y+0.3) () {21:00}; 
		\node[timeTick] at (8,\y+0.3) () {22:00};
		\node[timeTick] at (9,\y+0.3) () {23:00}; 
		\node[timeTick] at (10,\y+0.3) () {24:00};

		\node[probType,minimum width = 10cm, minimum height = 0.6cm, below=\y+0.3cm of startTime, anchor = north west] (da) {};		
		\node[probType,right=1pt of da.west,minimum width=9.95cm,fill=gray!50] {};
		\node[left=0.1cm of da.west,anchor=east,align=center] {DA-UC};
		\foreach \x in {0,1,2,3,4,6,7,8,9}{
			\node[circle,draw=red,scale=0.3,right=(\x) of da.west] {}; 
		}

		\node[probType,minimum width = 10cm, minimum height = 1.2cm, below=0.1cm of da] (st) {};
		\node[probType,below right=5pt and 1pt of st.north west,minimum width=4cm,fill=gray!50] (st_a) {};
		\foreach \x in {0,0.25,0.5,0.75,1,0,1.25,1.5,1.75,2.0,2.25,2.5,2.75,3.0,3.21,3.5,3.75}{
			\node[circle,draw=red,scale=0.3,right=(\x) of st_a.west] {}; 
		}	
		\node[probType,anchor=west,below right=5pt and 3cm of st_a.west,minimum width=4cm,fill=gray!50] (st_b) {};
		\foreach \x in {0,0.25,0.5,0.75,1,0,1.25,1.5,1.75,2.0,2.25,2.5,2.75,3.0,3.21,3.5,3.75}{
			\node[circle,draw=red,scale=0.3,right=(\x) of st_b.west] {}; 
		}	
		\node[probType,anchor=west,below right=10pt and 3cm of st_b.east,minimum width=4cm,anchor=east,fill=gray!50] (st_c) {};
		\foreach \x in {0,0.25,0.5,0.75,1,0,1.25,1.5,1.75,2.0,2.25,2.5,2.75,3.0,3.21,3.5,3.75}{
			\node[circle,draw=red,scale=0.3,right=(\x) of st_c.west] {}; 
		}	
		\node[left=0.1cm of st.west,anchor=east] {ST-UC};
		
		\path[thick,->,draw=gray!60] (da.west) edge[bend right] node[left] {} (st_a.west);
		\path[thick,->,draw=gray!60] ($(da.west) + (0:3)$) edge[bend right] node[left] {} (st_b.west);
		\path[thick,->,draw=gray!60] ($(da.west) + (0:6)$) edge[bend right] node[left] {} (st_c.west);

		\pgfmathsetmacro{\y}{-2.75}
		\draw[dashed] (0,\y) -- (1.8,\y); \draw[thick] (1.8,\y) -- (8.2,\y); \draw[dashed,->] (8.2,\y) -- (10,\y);
		\foreach \x in {2,4,6,8}{
			\draw[thick] (\x,\y-0.1) -- (\x,\y+0.1);		
		}
		\foreach \x in {2.5,3,3.5,4.5,5,5.5,6.5,7,7.5}{
			\draw (\x,\y-0.07) -- (\x,\y+0.07);
		}
		\node[timeTick,font=\footnotesize] at (2,\y+0.3) (edTime_a) {02:00}; 
		\node[timeTick,font=\footnotesize] at (4,\y+0.3) {03:00}; 
		\node[timeTick,font=\footnotesize] at (6,\y+0.3) {04:00}; 
		\node[timeTick,font=\tiny] at (2.5,\y+0.2) (edTime_b) {:15}; 
		\node[timeTick,font=\tiny] at (3.0,\y+0.2) (edTime_c) {:30}; 
		\node[timeTick,font=\tiny] at (3.5,\y+0.2) {:45}; 
		\node[timeTick,font=\tiny] at (4.5,\y+0.2) {:15}; 
		\node[timeTick,font=\tiny] at (5.0,\y+0.2) {:30}; 
		\node[timeTick,font=\tiny] at (5.5,\y+0.2) (edTime_d) {:45}; 
		\node[timeTick,font=\tiny] at (6.5,\y+0.2) {:15}; 
		\node[timeTick,font=\tiny] at (7.0,\y+0.2) {:30}; 
		\node[timeTick,font=\tiny] at (7.5,\y+0.2) {:45}; 
 	
		\node[probType,minimum width = 10cm, minimum height = 2cm, below=0.1cm of st] (rt) {};
		\node[probType,below=0.25cm of edTime_a,minimum width=2.5cm,fill=gray!50,anchor=north west] (rt_a) {};
		\foreach \x in {0,0.5,1,1.5,2}{
			\node[circle,draw=blue,scale=0.3,right=(\x) of rt_a.west] {}; 
		}
		\node[probType,below=0.5cm of edTime_b,minimum width=2.5cm,fill=gray!50,anchor=north west] (rt_b) {};
		\foreach \x in {0,0.5,1,1.5,2}{
			\node[circle,draw=blue,scale=0.3,right=(\x) of rt_b.west] {}; 
		}
		\node[probType,below=0.8cm of edTime_c,minimum width=2.5cm,fill=gray!50,anchor=north west] (rt_c) {};
		\foreach \x in {0,0.5,1,1.5,2}{
			\node[circle,draw=blue,scale=0.3,right=(\x) of rt_c.west] {}; 
		}
		\node[probType,below=1.2cm of edTime_d,minimum width=2.5cm,fill=gray!50,anchor=north west] (rt_d) {};
		\foreach \x in {0,0.5,1,1.5,2}{
			\node[circle,draw=blue,scale=0.3,right=(\x) of rt_d.west] {}; 
		}
		\draw[dotted,->] (rt_c) -- (rt_d.west);
		\node[left=0.1cm of rt.west,anchor=east] {HA-ED};
		
		\path[thick,->,draw=gray!60] ($(st_a.west) + (0:2)$) edge[bend right] node[left] {} (rt_a.west);
			
	\end{tikzpicture}
	\caption{Hierarchical structure of Operating Framework.} 	
    \label{fig:operatingFramework}
\end{figure*}

In this section, we introduce the details of the SHP framework illustrated in Fig. \ref{fig:operatingFramework}. We provide an abstract description of SP models used in each individual layer and defer the detailed description to Appendix A. In the models that constitute the DHP framework, the expectation-valued objective functions are replaced by their deterministic counterparts.

\subsection{The Multilayer Framework} \label{sec:decisionProcess:modelingsetup}

Our setup will capture operations of the power systems across multiple days. We use three subscripts, $[i]$, $[j]$, and $[k]$, corresponding to the DA-UC, ST-UC, and HA-ED models, respectively. Following the timescales presented in Table \ref{tab:modelParameters}, the subscript $[i]$ is used to index the twenty four (hourly) decision epochs of the DA-UC model instance corresponding to the $i$th day. Since a ST-UC instance is solved every three hours, eight ST-UC model instances are solved on a given day, each with a four-hour horizon. The subscript $[j]$ is used to index the sixteen ($15$-minute) decision epochs corresponding to the $j$th ST-UC instance. Finally, since a HA-ED instance is solved every $15$ minutes, ninety six HA-ED instances are solved on a given day. For the $k$th instance, the five decision epochs are captured by $[k]$. To make our notation clear, assume that the operations of the day begin at 12:00 am. The DA operations of the first day will be captured by DA-UC instance indexed by $ i = 1$; the ST operations between 9:00 am - 12:00 pm will be captured by the ST-UC instance indexed by $j = 4$; and the RT operations for 10:00 am will be captured by the HA-ED indexed by $k = 11$. 

Let ${\x}$ and ${\y}$ be the vector variables that model the generators' on/off statuses and production levels, respectively. We will use superscripts $d$, $s$, and $r$ on objective functions, constraint sets, and parameters associated with DA, ST, and HA, respectively. A subscript $[\bullet]$ (as in, $z_{[\bullet]}$) is used to refer to the decision vector corresponding the specific instances. The randomness associated with renewable supplies and electricity demand is captured by random vectors which evolve over time. We denote these random vectors by $\tilde{\xi}_{[i]}^d$, $\tilde{\xi}_{[j]}^s$, and $\tilde{\xi}_{[k]}^r$, for the $i^{th}$, $j^{th}$, and $k^{th}$ instances of the DA-UC, ST-UC, and HA-ED models, respectively. 

Any model instance is built using optimal solutions obtained by solving previous model instances at all three levels of the hierarchy. We summarize these optimal solutions as a state vector $\mathcal{H}_{[\bullet]}$. For instance, the $j$th instance of ST-UC is built using (i) the optimal solution of DA-UC instance corresponding to time periods $[j]$, (ii) the solutions of the previous ST-UC instances, viz. $[j-1], [j-2],$ etc., and (iii) the solutions of the HA-ED for time periods immediately preceding $[j]$. These will be collectively referred to as $\mathcal{H}_{[j]}$.

\begin{table}[h]
    \centering
	\begin{tabular}{c c c c}
		\hline 
		   & Horizon    & Resolution & Solution Frequency \\
		\hline
		DA & 24 hours     & 60 minutes & 24 hours    \\
		ST & 4 hours      & 15 minutes & 3 hours \\
		RT & 75 minutes & 15 minutes & 15 minutes \\
		\hline  
	\end{tabular}
    \caption{Model Timescale used for experiments} \label{tab:modelParameters}
\end{table}

Given the above notation, for a given day $i$, we define the DA-UC model as a two stage stochastic program. The first-stage involves commitment decisions and the second-stage has a security-constrained ED defined over the entire model horizon. This program can be abstracted as follows:
\begin{align}\label{eqn:dauc}
\text{DA-UC} \, \left( \mathcal{H}_{[i]}, \, \tilde{\xi}_{[i]}^d \right) = \min ~ f^d_{[i]}\left(\x_{[i]}, \y_{[i]}\right) \span \notag \\  
\text{subject to:} ~ \left({\x}_{[i]}, \, {\y}_{[i]}\right) \in \mathcal{X}^d_{[i]} \left( \mathcal{H}_{[i]}, \, \tilde{\xi}_{[i]}^d \right).
\end{align}
Above, the function $\text{DA-UC} \, (\cdot)$ uses the history of the generators (i.e., $\mathcal{H}_{[i]}$) and a representation of renewable supplies and demand stochastic process $\tilde{\xi}_{[i]}^d$ to determine commitment schedules for the DA generators. The feasible set of this  model is denoted as $\mathcal{X}^d_{[i]}\left( \cdot \right)$, and the expectation-valued function $f^d_{[i]}(\cdot)$ captures the combined commitment and dispatch costs. We denote the optimal solution of this model as $(\x^{\star,d}_{[i]}, \, \y^{\star,d}_{[i]})$. The optimal solution acts as an input to the lower level optimization problems, viz. ST-UC and HA-ED. Since the resolution of the lower level optimization problems is higher ($15$ minutes) when compared to the DA-UC ($60$ minutes), appropriate adjustments are made to keep time consistency. 

Using the optimal decisions obtained from previously solved model instances, i.e., $\mathcal{H}_{[j]}$, the $j$th instance of the two-stage stochastic program for ST-UC is given as follows:
\begin{subequations} \label{eqn:stuc}
\begin{align}
\text{ST-UC} \, \left(\mathcal{H}_{[j]} , \, \tilde{\xi}_{[j]}^s\right) = \min ~ f^s_{[j]} \left({\x}_{[j]}, \, {\y}_{[j]}\right) \span \notag \\  
\text{ subject to:} ~ & \left({\x}_{[j]}, {\y}_{[j]}\right) \in \mathcal{X}^s_{[j]} \left( \mathcal{H}_{[j]}, \, \tilde{\xi}_{[j]}^s \right), \notag \\ 
& {\x}_{[j]}^d = {\x}_{[j]}^{\star,d}, \label{eqn:stuc_fixedcommitments} \\
& |{\y}_{[j]}^d - {\y}_{[j]}^{\star,d}| \leq \epsilon_j. \label{eqn:stuc_generationDeviation}
\end{align}
\end{subequations}
The $\text{ST-UC} \, (\cdot)$ and $\text{DA-UC} \, (\cdot)$ are similar in nature, except for \eqref{eqn:stuc_fixedcommitments} and \eqref{eqn:stuc_generationDeviation}. The former constraint ensures that the DA commitment decisions are respected in the ST-UC model for generators that participate only in the DA market, and the latter allows for their generation levels to be updated only within a bound defined by the parameter $\epsilon_j$. Such bounds are placed to avoid myopic solutions of ST models as they have shorter horizons than the DA-UC model. \sa{The impact of alternative constraints, such as imposing only an end-state requirement that agrees with the generation amounts obtained in higher level planning, could also be investigated. In contrast, the decisions in our ST-UC model are restricted to be within $\epsilon_j$ of corresponding higher layer decisions across all time periods in the model horizon. Our choice for $\epsilon_j$ is to set them to generators' ramping limits. This ensures that generators can always ramp up/down to the production levels that DA-UC recommends, while still being able to deviate from them to a certain extent.}


Using all the commitment decisions and generation levels (captured by $\mathcal{H}_{[k]}$) prescribed by higher-level UC models, the $k$th instance of the two-stage HA-ED program is instantiated as shown below:
\begin{subequations} \label{eqn:haed}
\begin{align}
\text{ED} \, \left( \mathcal{H}_{[k]}, \tilde{\xi}_{[k]}^r \right) = \min ~ f^r_{[k]} \left({\x}_{[k]}, \, {\y}_{[k]}\right) \span \notag \\ \text{ subject to:} ~ & ({\x}_{[k]}, {\y}_{[k]}) \in \mathcal{X}_{[k]}^r \left( \mathcal{H}_{[k]}, \, \tilde{\xi}_{[k]}^r \right), \notag \\ & {\x}_{[k]}^d = {\x}^{\star,d}_{[k]},~ {\x}_{[k]}^s = {\x}^{\star,s}_{[k]} \label{eqn:ed-fixedcommitments} \\
& |{\y}_{[k]} - {\y}_{[k]}^{\star,s}| \leq \epsilon_k. \label{eqn:ed_generationDeviation}
\end{align}
\end{subequations}
The first-stage of HA-ED captures generation for the first decision epoch, whereas, the second-stage involves the remaining decision epochs in the planning horizon. Note that, only the first-stage decision (the here-and-now) decision is implemented. Since the dispatch decisions are fixed in \eqref{eqn:ed-fixedcommitments}, the resulting model only has continuous decision variables. As in the case of ST-UC, the constraint \eqref{eqn:ed_generationDeviation} ensures that the HA generation does not deviate beyond $\epsilon_k$ from the generation decision from the upper layer optimization problem (ST-UC, in this case). This is done to overcome the myopic nature of the HA-ED model resulting from its shorter horizon when compared to the UC models at higher levels of the hierarchy.

The deterministic variants of the models defined above use only point forecasts $\bar{\xi}_{[\cdot]}$ and the objective function is defined as the cost associated with both unit commitment decisions and dispatch under such forecast. The main distinction between the models used in the DHP and SHP hierarchies is that the objective functions of the latter are defined with a deterministic first-stage cost and an expected recourse (second-stage) cost. 
The recourse cost is the optimal objective of a second-stage optimization model that is instantiated by the first-stage decisions $x$ and a realization of the random variable $\tilde{\xi}$. 

\subsection{Modeling and Solving Individual Optimization Problems} \label{sec:decisionProcess:modelingandsolving}
The deterministic UC and ED models are solved using MIP and LP algorithms available in off-the-shelf solvers (e.g., CPLEX, Gurobi). In the 2-SP models for UC, we use a finite set of scenarios to represent the uncertainty (100 scenarios, to be specific). Even with modest numbers of scenarios, the resulting deterministic equivalent models could be very large and cannot be handled by off-the-shelf solvers. Given the recourse problems are LPs and can be decoupled by scenarios, we use the well-known L-shaped method (also known as Benders decomposition) to solve the stochastic UC problems \citep{Benders1962, VanSlyke1969}. The 2-SP formulation of HA-ED problem has LPs in both stages. We solve these models using a sequential sampling method called regularized stochastic decomposition (SD) algorithm \citep{Higle1994}. Table \ref{tab:modelsAndAlgorithms} provides a summary of model characteristics and the adopted algorithmic choices. 

\begin{table}[ht]
    \centering
    \small
    \resizebox{0.9\textwidth}{!}{
    \begin{tabular}{|c|c c|c c|} \hline
    Operational & \multicolumn{2}{c}{Deterministic} & \multicolumn{2}{|c|}{Stochastic} \\ \cline{2-5}
    Layer & Problem type & Solver & Problem type & Algorithms \\ \hline
    DA-UC & MIP & CPLEX default & Stochastic MIP & Benders/L-Shaped \\
    ST-UC & MIP & CPLEX default & Stochastic MIP & Benders/L-Shaped \\
    HA-ED & LP & CPLEX default & Stochastic LP & Regularized SD \\ \hline
    \end{tabular}}
    \caption{Problem class and solution algorithms used for individual instances}
    \label{tab:modelsAndAlgorithms}
\end{table}

\sa{
The solution quality of an SP is directly influenced by the number of scenarios considered. We label a solution to an SP model to be \textit{statistically-optimal} (for a given confidence level $\alpha$) if the optimal value reported by the algorithm falls within an acceptable confidence interval computed using an independent validation process. We refer the reader to \cite{Sen2016} for details regarding statistical optimality and the validation process. In our setup, the SD algorithm can achieve statistical optimality by increasing its sample size on the fly, whereas a fixed sample size used in stochastic mixed-integer UC models makes statistical optimality is too demanding a requirement to impose for the current state-of-the-art.
}

In our framework, our knowledge of the underlying stochastic process improves as we get closer to the time of dispatch. This leads to more accurate point forecasts (from DA to ST, and eventually HA forecasts) of wind and solar generation amounts. Deterministic models are set up using the point-forecast time series available just before their execution. On the other hand, the scenarios used to instantiate and solve stochastic UC and ED problems are generated using separate time series simulators for solar and wind. The simulators are based on a vector autoregressive (VAR) model that captures the spatio-temporal correlation of the stochastic processes governing generator outputs. The VAR models are trained using the point-forecast time series available immediately before the execution of the respective stochastic models. We refer the reader to Appendix B for an extensive discussion on the prediction and scenario generation aspects of our study.

\subsection{Evaluating the Performance of Hierarchical Planning Frameworks} \label{sec:decisionProcess:evaluation}
\sa{
We evaluate the performance of both DHP and SHP frameworks by solving the individual model instances in a rolling horizon fashion over multiple days against the historical observations of demand and renewable generation (see Fig.\ \ref{fig:operatingFramework}). The historical commitment decisions, summarized in $\mathcal{H}_{[i]}$ with $i = 1$, serve as input to our simulation setup. The DA-UC and ST-UC instances use a representation of the stochastic process that is built based on the forecast data. On the other hand, the first-stage of HA-ED uses the actual observations, while the scenarios in the second-stage are based on forecast available at the time when the model is instantiated. Once the HA-ED problem is solved, the generation amounts are recorded and all metrics (such as load shedding, cost) are computed based on these values. The simulation clock is then moved forward and the next problem (possibly from the upstream layers of the hierarchy) is solved. Notice that only the HA-ED problem's first-stage generation solutions (and the associated commitment decisions) constitute our final outputs, while the solutions of the UC problems can only \emph{influence} these outcomes.
}



 
\section{Experimental Study} \label{sec:experiments}
The NREL-118 instance includes 327 generators (75 solar and 17 wind), 118 buses, and 186 transmission lines, along with DA forecasts and real-time outputs of renewable generators and demand. This dataset has a power system that is rich in solar, wind, and hydro resources, and may be considered as forward-looking. We assess three factors that have a significant impact on power system operations. These are solar and wind integration, reserve requirements, and the planning strategy adopted for handling the UC and ED problems.  By varying these factors, we analyze their impact on certain reliability metrics, as well as economic and environmental ones, such as unmet demand, operating costs, and GHG emissions.

In Table \ref{tab:factor_levels}, we summarize the values considered for these factors in our experiments. We use a three-letter code to identify specific types of models (i.e., \textbf{D}eterministic or \textbf{S}tochastic) used within the three-layer planning hierarchy (DA, ST, RT). Following this notation, the DDD setting is the benchmark deterministic planning framework (i.e., DHP), whereas both DDS and SDS can be considered as examples of the SHP framework. 

\sa{Note that in actual operations, both the ST-UC and HA-ED problems are solved during the dispatch-day and optimal solutions from both must be computed in very short time-intervals. This is especially important for ST-UC, as these problems commit fast-ramp resources to respond to reliability concerns which may emerge in real time. For the HA-ED solution obtained from stochastic linear programs, the state-of-the-art algorithms are capable of delivering \textit{statistically-optimal solutions rapidly} using very detailed representations of the stochastic processes. For instance, in our experiments the SD algorithm that solves the HA-ED instances used a minimum of $256$ simulated multivariate time series of wind and solar stochastic processes ($75$ solar and $17$ wind locations resulting in a $92$ dimensional time series). However, the solver technology for stochastic mixed-integer programs is nowhere near the same level of maturity to deliver high quality solutions within tight computational time bounds. On these grounds, academic studies on these models are typically conducted with only a handful of scenarios, and their solutions are seldom validated to ensure that the coarse representation of uncertainty indeed provides a prediction of the optimal objective function which lies within an upper bound confidence interval of the true SP problem. In contrast, our study adopts the prudent step of following the standard industry practice today, namely, solving ST-UC problems as deterministic mixed-integer programs. In subsequent sections, our experiments will reveal promising results even with a deterministic ST-UC, providing \textit{lower bounds on achievable gains} through the SHP framework with fully-stochastic layers\footnote{For a given input state, the stochastic ST-UC will outperform the deterministic ST-UC due to the well known non-negativity of the value of stochastic solution (VSS) \citep{Birge1997}.}. This also reinforces the need for further research on two aspects of renewable integration: a) reasonably accurate probabilistic forecasts of wind energy over short intervals of time \citep[less than six hours;][]{carroll2018high} and b) specialized (fast) algorithms for large scale ST-UC problems which can deliver near-optimal solutions for real-scale models with binary (start-up/shut-down) variables. For this reason, our study does not include the following combinations: ($\bullet$, S, $\bullet$).}

\begin{table}[ht]
    \small 
    \centering
    \resizebox{0.99\textwidth}{!}{
    \begin{tabular}{lp{1.5cm}p{8cm}}
        \toprule
        \textbf{Category} & \textbf{Label} & \textbf{Description} \\ 
        \midrule
        \multirow{3}{1.8cm}{\textbf{Solar \& Wind Integration}}
        & \multirow{1}{*}{Low SW:} & Original solar \& wind outputs in the NREL-118 dataset. \\ 
        & Med.\ SW: & Twice the original values. \\ 
        & High SW: & Thrice the original values. \\ 
        \midrule
        \multirow{4}{2cm}{\textbf{Reserve  \\ Requirements}}
        & Very Low: & 5\% for UCs, 1.25\% for ED. \\ 
        & Low: & 10\% for UCs, 2.5\% for ED. \\ 
        & Med.\: & 15\% for UCs, 5\% for ED. \\ 
        & High: & 20\% for UCs, 10\% for ED. \\ 
        \midrule
        \multirow{3}{2cm}{\textbf{Planning Setting}}
        & DDD: & Deterministic DA-UC, ST-UC, ED. \\ 
        & DDS: & Deterministic DA-UC, ST-UC; stochastic ED. \\ 
        & SDS: & Deterministic\ ST-UC; stochastic DA-UC, ED. \\ 
        \bottomrule
    \end{tabular}}
    \caption{Investigated factors and their levels.}
    \label{tab:factor_levels}
\end{table}

\sa{In all experiments, we consider a total planning horizon of 7 days, and a maximum resolution of 15 minutes. Recall that evaluations of the hierarchical frameworks are carried out in a rolling horizon manner. We assume that the DA-UC and ST-UC problems commit distinct sets of generators, and ST-UC problems cannot decommit DA resources (see Appendix A for details). The reported metrics are based on generator heat rates, fuel consumption rates, and emission rates given in the NREL-118 dataset, as well as first-period generation decisions of HA-ED problems that are based on actual observations. Note that both the deterministic and the stochastic models are evaluated based on the same actual observations time series. In what follows, we present the results from these evaluations.
}

\subsection{Reliability Impact} \label{sec:experiments:reliability}
Significant amounts of unmet demand revealed in the planning process may result in blackouts with damaging economic consequences for system constituents. Due to its importance to ISOs as well as customers, we start our discussion by presenting the average and maximum unmet demand values in Table \ref{tab:unmet_demand}.

\begin{table}[ht]
    \centering
    \small
    \begin{tabular}{l|l||r|r|r||r|r|r}
    \toprule
    \multicolumn{2}{c||}{ } & \multicolumn{3}{|c||}{Avg.\ Unmet Demand} & \multicolumn{3}{|c}{Max.\  Unmet Demand} \\
    \midrule
    Planning & Reserve & \multicolumn{3}{c||}{Solar \& Wind Integ.\ } & \multicolumn{3}{c}{Solar \& Wind Integ.\ } \\
    Setting & Req.\ & Low & Med.\ & High & Low & Med.\ & High \\ 
    \midrule
DDD & V Low & 1.9 & 4.4 & 17.0 & 320.0 & 595.1 & 956.1 \\
 & Low & 0.5 & 3.0 & 1.5 & 246.9 & 336.5 & 274.3 \\
 & Med. & 0.0 & 0.4 & 1.2 & 0.0 & 252.6 & 552.5 \\
 & High & 0.0 & 0.0 & 0.0 & 0.0 & 0.0 & 0.0 \\
DDS & V Low & 1.2 & 0.7 & 3.9 & 172.0 & 162.8 & 494.3 \\
 & Low & 0.0 & 0.0 & 0.5 & 0.0 & 0.8 & 217.8 \\
 & Med. & 0.0 & 0.0 & 0.0 & 0.0 & 0.0 & 0.0 \\
 & High & 0.0 & 0.0 & 0.0 & 0.0 & 0.0 & 0.0 \\
SDS & V Low & 0.0 & 1.9 & 0.6 & 19.2 & 182.0 & 356.6 \\
 & Low & 0.0 & 0.0 & 0.0 & 0.0 & 0.0 & 0.0 \\
 & Med. & 0.0 & 0.0 & 0.0 & 0.0 & 0.0 & 0.0 \\
 & High & 0.0 & 0.0 & 0.0 & 0.0 & 0.0 & 0.0 \\
 \bottomrule
 \end{tabular}
    \caption{Average and maximum unmet demand amounts (MW)} \label{tab:unmet_demand}
\end{table}

In general, we notice higher unmet demand when more solar and wind resources are introduced to the system with fully deterministic planning (DDD). More conservative reserve requirements substantially reduce these values, but possibly comes with an additional monetary cost. \textit{On the other hand, adopting stochastic planning models into the modeling framework can zero out unmet demand, even at less conservative reserve requirements}. For instance, to completely eliminate unmet demand from the planning process, DDD, DDS, and SDS necessitate high, medium, and low reserve requirements, respectively, under Medium SW and High SW settings. This observation supports the use of stochastic planning models to accommodate the uncertainty associated with VERs and reduce reliance on (manually-imposed) reserve restrictions. More importantly, it suggests the possibility of a more economic way of operating the system.

During our planning process, certain fast-ramping generators can be committed by the ST-UC problems to recover from unexpected supply shortages during the day. We evaluate the reliance on ST-UC problems by looking at the percentage of time that these generators were active (see Table \ref{sec:experiments:tab:reliance_on_stuc}). We observe a consistent trend where the DDD setting heavily relies on ST-UC problems to maintain reliability. In contrast, DDS and SDS settings substantially reduce these requirements even under higher renewable-integration settings.

\begin{table}[ht]
\centering
\small
\begin{tabular}{c|rrrr}
\toprule
\multicolumn{1}{c|}{Planning Setting} & Low SW & Med.\  SW & High SW\\ 
\midrule
{DDD} & 11.0 & 11.7 & 11.8 \\ 
{DDS} & 9.7 & 8.3 & 7.8 \\ 
{SDS} & 8.7 & 6.5 & 5.3 \\ 
\bottomrule
\end{tabular}
\caption{Average percentage of time ST-UC generators were active.}\label{sec:experiments:tab:reliance_on_stuc}
\end{table}

\begin{figure}[ht]
\centering
\includegraphics[trim={5mm 1mm 2mm 1mm}, clip, width=0.77\columnwidth]{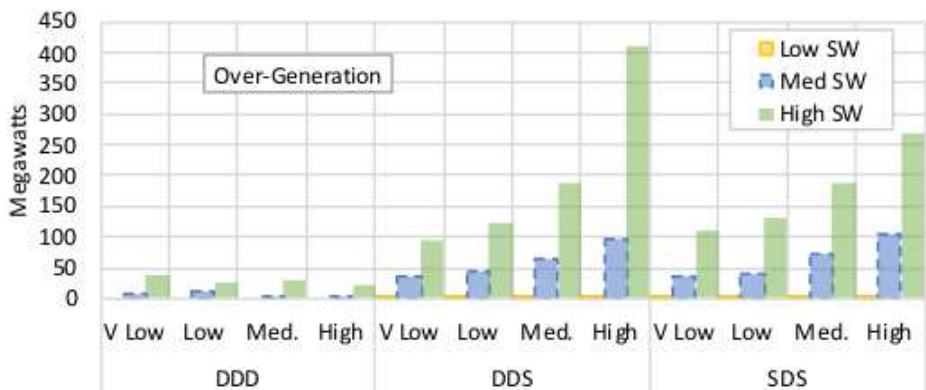}
\caption{Average over-generation amounts by conventional generators.}
\label{sec:experiments:fig:over_generation}
\end{figure}

Fig.\ \ref{sec:experiments:fig:over_generation} shows the average amounts of over-generation (by conventional generators) estimated under all settings. \sa{Recall that in the absence of significant storage capacity in the system (such as today), vast swings in renewable supply may result in over-generation by conventional generators \citep{Olson2015}. This, in turn, leads to negative electricity prices or costly exchanges between neighboring ISOs to ensure supply and demand matches without damaging physical infrastructure or posing financial concerns for market participants \citep{Penn2017}.} 

We observe the smallest over-generation amounts under the DDD setting. This is not surprising as it would never be optimal to over-produce in a deterministic optimization model provided that ramping capabilities are sufficient to cover ramping needs within the model's horizon. In contrast, stochastic optimization (i.e., DDS and SDS) compensates for the variability in future time periods by over-generating in significantly larger amounts, thereby preventing situations where upwards-ramping capabilities may not be sufficient under certain settings.  This may actually mitigate the Duck-Chart phenomenon discussed in the introduction.

\begin{figure}[ht]
\centering
\includegraphics[trim={5mm 1mm 2mm 1mm}, clip, width=0.77\columnwidth]{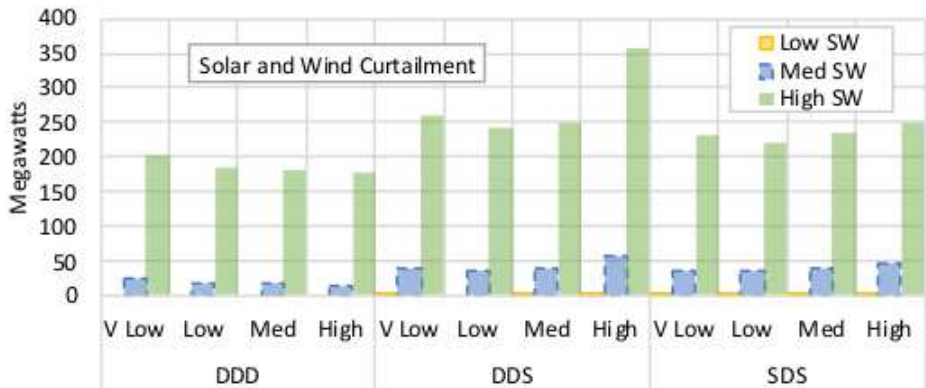}
\caption{Average curtailed solar and wind energy.}
\label{sec:experiments:fig:curtailment}
\end{figure}

In terms of solar and wind curtailment, Fig.\ \ref{sec:experiments:fig:curtailment} shows a significant trend where higher renewable integration leads to substantial amounts of curtailment, providing support for the need to introduce greater energy storage. In addition, we still observe that both DDS and SDS lead to slightly more curtailment than that in DDD.

\sa{
Overall, the outcomes we observed so far suggests that a transition to SHP leads to reduced loss of load, through conventional over-generation and renewable curtailment that is dynamically adjusted across the buses based on the uncertainty lying ahead. This indicates that reliability can be improved by better planning models, as opposed to increasing the reserve requirements. In subsequent sections, we will see promising results on the economic and environmental impacts of operating the network with lower reserve requirements, which will reveal the contributions of stochastic modeling for grids with high renewable integration.
}

\begin{figure}[ht]
    \centering
    \ifpreprint
        \includegraphics[width=0.7\textwidth]{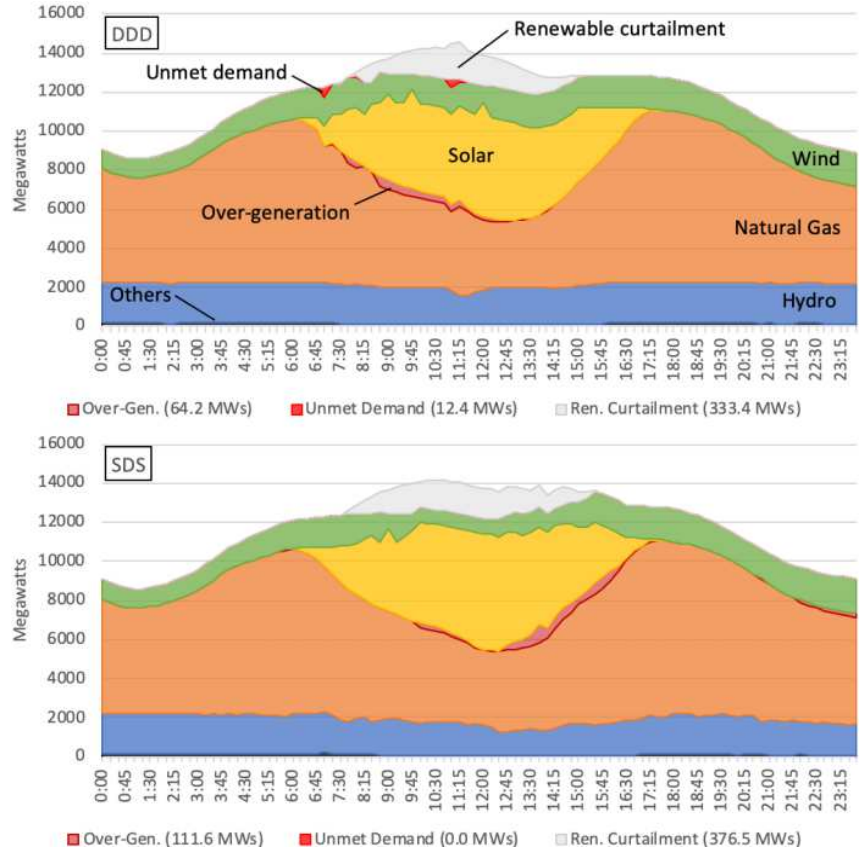}
    \else
    \begin{subfigure}[t]{0.49\textwidth}
        \centering
        \includegraphics[width=0.48\textwidth]{figures/generation_mix_2_1}
    \end{subfigure}
    \begin{subfigure}[t]{0.49\textwidth}
        \centering
        \includegraphics[width=0.48\textwidth]{figures/generation_mix_2_2}
    \end{subfigure}
    \fi
\caption{Generation mix, unmet-demand, and over-generation in DDD and SDS settings (very low reserves, High SW, and a sample day).}\label{sec:experiments:fig:generation_mix}
\end{figure}

Fig.\ \ref{sec:experiments:fig:generation_mix} illustrates intra-day generation profiles under DDD and SDS settings. Notice that the duck-chart is clearly visible. Another phenomenon to notice is that unmet demand, over-generation, and renewable curtailment may all occur simultaneously (at different buses), with the first occurring typically at day-time, when solar generators are active, and transmission capacity constitutes the bottleneck.  This underscores the importance of accounting for the transmission networks in power system experiments. As seen from the figure under discussion, SDS leads to more over-generation but reduces unmet demand from 16.9 MW to 0.7 MW. Furthermore, we observe higher variability in hydro-based generation under SDS (coefficient of variation of hydro-based generation is 0.13 in SDS vs.\ 0.05 in DDD). Hydro generators have better ramping capabilities, making them better suited to accommodate uncertainty. SDS naturally leverages this fact seamlessly.  

\subsection{Economic Impact} 
\label{sec:experiments:economics}
\begin{figure}[ht]
\centering
\includegraphics[width=0.97\columnwidth]{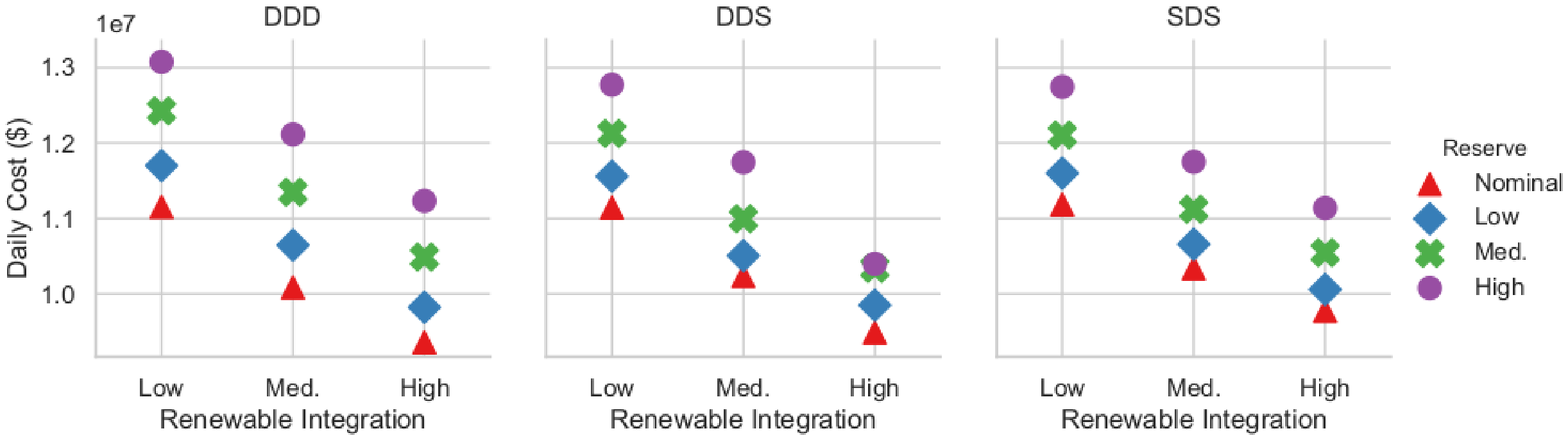}
\caption{Average daily generation cost of the power system under different reserve requirements and operations planning strategies.}
\label{sec:experiments:fig:avg_daily_cost}
\end{figure}

Fig.\ \ref{sec:experiments:fig:avg_daily_cost} demonstrates the average daily operating costs recorded in our experiments. The reported amounts exclude the potential costs associated with the consequences of operating a grid with low reliability (e.g., penalties associated with load-shedding). In line with expectations, increased renewable integration leads to lower costs \sa{as these resources have negligible generation costs. On the other hand, increased reserve requirements have the opposite effect, since more (and possibly expensive) resources need to be committed to maintain these requirements. Table \ref{sec:experiments:tab:avg_num_generators} illustrates this, where there is a consistent trend of growing number of committed generators as the reserve requirements are increased.
}

\begin{table}[ht]
\small
    \centering
    \begin{tabular}{l|c|c|c|c}
    \toprule
	\multicolumn{1}{l}{Reserve Req:} & \multicolumn{1}{c}{V Low} & \multicolumn{1}{l}{Low} & \multicolumn{1}{l}{Med.} & \multicolumn{1}{l}{High} \\ 
	\midrule
 DDD & 162.9 & 170.7 & 180.3 & 186.4 \\
 DDS & 157.6 & 162.8 & 168.0 & 174.2 \\ 
 SDS & 154.5 & 159.8 & 166.3 & 174.6 \\
 \bottomrule
 \end{tabular} 
 \caption{Average number of committed generators in each framework across different reserve requirement levels.} \label{sec:experiments:tab:avg_num_generators}
\end{table}

\sa{
Table \ref{sec:experiments:tab:avg_num_generators} also reveals another trend. The number of committed generators under DDS and SDS settings are consistently lower than that of DDD. A particular reason for this is the number of ST generators that are committed near dispatch time, which has already been illustrated in Table \ref{sec:experiments:tab:reliance_on_stuc}. Overall, however, we posit that the stochastic models can mitigate concerns over reliability without committing an abundance of resources.}

We now turn our attention to cases where reliability is achieved with the smallest reserves. Specifically, we look at reserve requirement levels at which the network demand is seamlessly fulfilled. Table \ref{sec:experiments:tab:avg_daily_cost_demand_met}
presents the daily operating costs corresponding to the minimum reserve requirements that must be set at each level of the hierarchy in order to ensure zero unmet demand. The cost figures indicate that, with stochastic optimization, reserve requirements can be relaxed as the models are able to dynamically adjust production levels by accounting for uncertainty in the future. As a result, the operating cost of the network can be reduced by up to $10.4\%$ (e.g., compare $\$11.23$M with DDD to $\$10.06$M with SDS, under high renewable integration). This is a promising finding as it mitigates financial concerns regarding the use of higher shares of VER in the grid \cite[e.g.,][]{Zappa2019}.

\begin{table}[ht]
\small
    \centering
    \begin{tabular}{l|c|c|c}
    \toprule
	& DDD	& DDS &	SDS \\ 
	\midrule
 Low SW & 12.42 (Med.) & 11.56 (Low) &  11.60 (Low) \\
 Med.\ SW & 12.11 (High) & 11.00 (Med.) & 10.66 (Low)  \\ 
 High SW & 11.23 (High) & 10.34 (Med.) & 10.06 (Low) \\
 \bottomrule
 \end{tabular} 
 \caption{Avg.\ daily operating cost of the system corresponding to the minimum reserve requirements leading to zero unmet demand (in million \$; Reserve requirements in parenthesis).} \label{sec:experiments:tab:avg_daily_cost_demand_met}
\end{table}

\sa{
For reserve requirements higher than those cited in Table \ref{sec:experiments:tab:avg_daily_cost_demand_met}, conflicting outcomes can be observed. For instance, the operating costs of DDS is better than that of SDS under medium and high reserve requirements, and high renewable integration. Clearly, this would not be a suggested manner of operations, as it costs more compared to what Table \ref{sec:experiments:tab:avg_daily_cost_demand_met} reports. However, it shows that just the stochastic HA-ED can accommodate fluctuations in VERs well with high-enough reserve requirements. Specifically, the deterministic DA-UC provides a (short-sighted) cost-optimal generator mix for a given forecast, and stochastic HA-ED ensures that not many additional ST generators are necessary during dispatch time (see Table \ref{sec:experiments:tab:reliance_on_stuc}). In comparison, the stochastic DA-UC provides a generator mix that is more conservative, therefore ready for scenarios that can jeopardize reliability, ensuring the sufficiency of lower reserve requirements.
}

\subsection{Environmental Impact} \label{sec:experiments:emissions}
\begin{figure}[ht]
\centering
\includegraphics[trim={0mm 3mm 12mm 5mm}, clip, width=0.87\columnwidth]{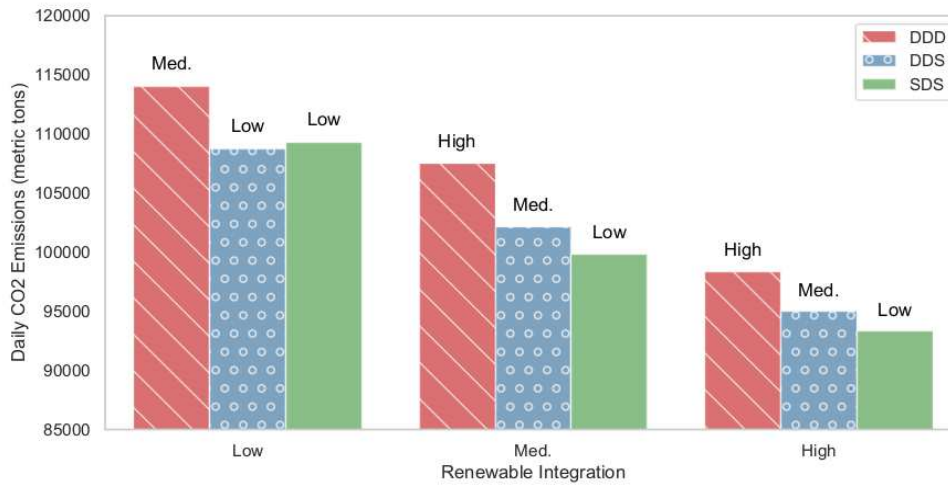}
\caption{Average daily CO$_2$ emissions of the power system corresponding to the minimum reserve requirements leading to zero unmet demand (reserve requirements are noted on top of the bars).}
\label{sec:experiments:fig:avg_daily_co2_demand_met}
\end{figure}

To assess the environmental impact of increasing renewable energy in the power system, we estimate the daily GHG emissions using the recorded generation amounts and mixes. Analogous to Table \ref{sec:experiments:tab:avg_daily_cost_demand_met}, 
Fig.\ \ref{sec:experiments:fig:avg_daily_co2_demand_met} demonstrates daily CO$_2$ emission estimates under the minimum reserve requirements that lead to zero unmet demand. These estimates are based on generators' heat and emission rates, which are given in the NREL-118 dataset, as well as their generation levels, which is determined by the optimization. Similar observations were made for the NO$_\text{x}$ and SO$_2$ emissions.

Concerning CO$_2$ emissions, we have two observations. First, in the experimented power system, higher renewable integration leads to lower levels of CO$_2$ emissions. While this sounds intuitive, opponents of this intuition typically suggest that the duck-chart phenomenon could lead to more emissions. This increase is attributed to over-generation and reliance on significant amount of gas-fired fast generators to overcome insufficient ramping capabilities and volatility of VERs. Our experiments suggest that this is not the case for power systems with similar characteristics. Second, while stochastic modeling (i.e., DDS, SDS) also leads to over-generation and renewable curtailment (see Fig.\ \ref{sec:experiments:fig:over_generation}-\ref{sec:experiments:fig:curtailment}), their impact can largely be reversed by the lower reserve requirements necessary to achieve the same level of reliability. In this regard, a concurrent optimization-simulation approach to obtain statistically appropriate measure of reserve requirement is presented in \cite{wang2020statistical}.

\sa{
\section{Conclusions and Discussion} \label{sec:conclude}
}
We presented the SHP framework for power systems with large-scale VER integration. While the call for a framework comprising of stochastic dynamic problems evolving at different timescales has been made before \citep[e.g.,][]{Ilic2007}, this is the first study to conduct comprehensive computational experiments with such a framework under a centralized planner. Our framework captures the operations and their interactions across day-ahead, short-term, and hour-ahead timescales. 

The optimization models in the SHP framework are solved using stochastic programming (SP) algorithms, some of which have been studied rigorously for over thirty years \citep{Higle1991, Higle1994} with more recent versions in \cite{Sen2016, Sen2014a}. In this paper, we coalesce our research from SP, including stochastic mixed-integer programming \citep{Atakan2018phbab}, with the work in power systems research for economic dispatch (ED) \citep{Gangammanavar2015, Gangammanavar2018} and unit commitment (UC) \citep{Atakan2018uc}. 

Our experiments indicate that the SHP framework overcomes many of the shortcomings of the DHP approach currently in practice. We observed that the SHP framework typically outperforms DHP in terms of reliability: Even at lower reserve requirements levels, SHP is more effective in eliminating unmet demand. Moreover, reliance on ST-UC problems (to avoid unmet demand) reduces by using SHP. On the other hand, the SHP framework leads to more conventional over-generation and renewable curtailment. The introduction of utility-scale storage devices can mitigate both of these drawbacks. The considered NREL-118 dataset is a reasonable approximation of networks with high renewable integration. Remarkably, our study provides hope that the grid of the future may be able to operate \textit{reliably} at lower levels of reserve requirements, while simultaneously reducing both operating costs, as well as GHG emissions. While our paper has made a strong case for a transition to SHP, we have only begun the discussion, and there is much to be gained by refining our methods to incorporate sophisticated SMIP algorithms \citep[e.g.,][]{Gade2014, Yuan2009}. These algorithms are better suited to include start-up/shut-down variables in the second-stage of stochastic UC models, such as those necessary for advanced ST-UC modeling. We suspect that specialized versions of these algorithms may bear fruit in future studies of the SHP framework. 

Given the current climate-change concerns, (i) efficient generator designs and power electronics, (ii) market design, and (iii) optimization software used in planning and operations can help address the challenges associated with meeting ambitious renewable portfolio standards. To ``tame the duck'', advances along all three fronts will be critical. Efforts in this paper lay the groundwork for addressing (iii) through the SHP framework and the use of stochastic optimization tools. 
These results point to the next steps that should involve experiments with actual ISO data. We hope to undertake this research as part of our future endeavors.

\section*{Acknowledgments}
This research was supported by the National Science Foundation (Grant No: CMMI 1822327). We thank the referees for their thorough reading, their recommendations, and even seeking clarifications which helped bring greater focus on aspects which are likely to promote greater renewable-energy integration.
\section{Detailed Mathematical Models} \label{sec:models}

\newcommand{\usedgen}{G^+}
\newcommand{\overgen}{G^-}

\setlength{\columnsep}{0.4cm}
\setlength{\columnseprule}{0.2pt}
\begin{table} [h!]
\centering
\footnotesize
\begin{tabular}{c}
\toprule
\textbf{Sets} \\ 
\midrule
\begin{minipage}{0.99\textwidth}
\begin{multicols}{2}
\begin{description}
\item[$\mathcal{B}$:] buses.
\item[$\mathcal{L}$:] transmission lines.
\item[$\mathcal{G}$:] generators.\
\item[$\mathcal{G}_r$:] solar and wind generators.
\item[$\mathcal{G}_c$:] conventional generators ($\mathcal{G}_c = \mathcal{G} \setminus \mathcal{G}_r$).
\item[$\mathcal{G}_j$:] generators located in bus $j \in \mathcal{B}$.
\item[$\mathcal{G}^{DA}$:] Slow-ramp DA generators.
\item[$\mathcal{G}^{ST}$:] Fast-ramp ST generators.
\item[$[t{]}$:] time periods.
\end{description}
\end{multicols}
\end{minipage} \smallskip \\ 
\midrule
\textbf{Parameters} \\ 
\midrule
\begin{minipage}{0.99\textwidth}
\begin{multicols}{2} 
\begin{description}
\item[$G^{\max}_g$:] generation capacity of $g \in \mathcal{G}_c$.
\item[$G^{\min}_g$:] minimum generation requirement for $g \in \mathcal{G}_c$.
\item[$\Delta G^{\max}_g$:] ramp up limit for $g \in \mathcal{G}_c$.
\item[$\Delta G^{\min}_g$:] ramp down limit for $g \in \mathcal{G}_c$.
\item[$UT_g$:] minimum uptime requirement of $g \in \mathcal{G}_c$.
\item[$DT_g$:] minimum downtime requirement of $g \in \mathcal{G}_c$.
\item[$B_{ij}$:] susceptance of arc $(i,j) \in \mathcal{L}$.
\item[$D_{jt}$:] load in bus $j \in \mathcal{B}$, in period $t \in [t]$.
\item[$R_{jt}$:] reserve requirement in bus $j \in \mathcal{B}$ and period $t \in [t]$.
\item[$F_{ij}^{\max}$:] maximum permitted flow through arc $(i,j) \in \mathcal{L}$.
\item[$c^{s}_g$:] start up cost of $g \in \mathcal{G}$.
\item[$c^{f}_g$:] no-load cost of $g \in \mathcal{G}$ (i.e., the intercept of the cost curve).
\item[$c^{p}_g$:] variable generation cost of $g \in \mathcal{G}$ (i.e., the slope of the cost curve).
\item[$\theta^{\max}_j$:] upper bound on the voltage-angle at bus $j \in \mathcal{B}$.
\item[$\theta^{\min}_j$:] lower bound on the voltage-angle at bus $j \in \mathcal{B}$.
\item[$\phi^o_g$:] penalty for over-generation by $g \in \mathcal{G}_c$.
\item[$\phi^c_g$:] penalty for renewable curtailment in $g \in \mathcal{G}_r$.
\item[$\phi^u_j$:] penalty for unmet demand in bus $j \in \mathcal{B}$. 
\end{description}
\end{multicols} 
\end{minipage} \smallskip \\ 
\midrule
\textbf{Decision Variables} \\ 
\midrule
\begin{minipage}{0.99\textwidth}
\begin{multicols}{2} 
\begin{description}
\item[$s_{gt}$:] 1 if $g \in \mathcal{G}$ is turned on in $t \in [t]$, 0 otherwise.
\item[$x_{gt}$:] 1 if $g \in \mathcal{G}$ is operational in $t \in [t]$, 0 otherwise.
\item[$z_{gt}$:] 1 if $g \in \mathcal{G}$ is turned off in $t \in [t]$, 0 otherwise.
\item[$\usedgen_{gt}$:] generation amount of $g \in \mathcal{G}$, in $t \in [t]$, which is consumed by the grid.
\item[$\overgen_{gt}$:] over-generation amount by $g \in \mathcal{G}_c$, in $t \in [t]$.
\item[$\overgen_{gt}$:] renewable curtailment in $g \in \mathcal{G}_r$, in $t \in [t]$.
\item[$F_{ij, t}$:] electricity flow through $(i, j) \in \mathcal{L}$, in $t \in [t]$.
\item[$\theta_{jt}$:] voltage angle at $j \in \mathcal{B}$, in $t \in [t]$.
\item[$D^{\text{shed}}_{jt}$:] amount of unmet load at $j \in \mathcal{B}$, in $t \in [t]$.
\end{description}
\end{multicols} 
\end{minipage} ~ \vspace{2mm} \\ 
\bottomrule
\end{tabular}
\caption{Nomenclature for mathematical formulations.} \label{tab:nomenclature}
\end{table}

In this section we outline the mathematical models of DA-UC, ST-UC, and HA-ED problems. We first provide the definition of the decision variables and describe the constraints, then, refer to these while constructing individual models. All models are presented for an arbitrary planning interval. Our notation is summarized in Table \ref{tab:nomenclature}. Recall that $[t]$ denotes the decision epochs of the problem being considered. 

We denote by $\mathcal{G}$ the set of all generators in the system. This set includes the both the conventional generators $\mathcal{G}_c$ and renewable generators $\mathcal{G}_r$. Furthermore, we assume that only a subset of conventional generators (typically gas-fired) are capable of providing fast-start services. We denote this set of generators by $\mathcal{G}^{ST}_c \subset \mathcal{G}_c$. Therefore, in our setting we commit a subset $\mathcal{G}_c^{DA} \subset \mathcal{G}_c$ in the DA-UC and the remaining generators are committed in the ST-UC. We assume that ST-UC cannot decommit DA resources. We also assume that renewable generators are always committed, however, their production levels can be curtailed. 

Generator commitment decisions are often modeled using three sets of binary variables $(x_{gt}$, $s_{gt}$, $z_{gt})$ that indicate whether $g$ is operational, turned on, and turned off in period $t$, respectively \citep{Garver1962}. These variables are linked with the following constraints\footnote{$\mathcal{G}_c^{*}$ refers to $\mathcal{G}_c^{DA}$ or $\mathcal{G}_c^{ST}$ based on whether the constraint appears in DA-UC and ST-UC models, respectively.}:
\begin{equation}
\label{con:state_equations}
    x_{gt} - x_{gt-1}  = s_{gt} - z_{gt}, \qquad \forall g \in \mathcal{G}_c^{*}, \, t \in [t].
\end{equation}

To model the minimum uptime and downtime requirements of generators, we use the turn on and off inequalities of \cite{Rajan2005}:
\begin{align}
\label{con:minimum_up_requirements}
    \sum_{j=t-UT_g+1}^{t-1} s_{gt} \leq x_{gt}, \qquad \forall g \in \mathcal{G}_c^*,\, t \in [t], \\ 
\label{con:minimum_down_requirements}    
    \sum_{j=t-DT_g}^{t} s_{gt} \leq 1-x_{gt}, \qquad \forall g \in \mathcal{G}_c^*, \, t \in [t].
\end{align}

We use two sets of variables to model generation levels. The $\usedgen_{gt}$ variable denotes the amount of electricity produced by generator $g$ in period $t$ \emph{and consumed by the grid}. The second variable $\overgen_{gt}$ can assume two different meanings, depending on the type of the generator. For solar and wind generators (i.e., $\forall g \in \mathcal{G}_r$), these variables capture the amount of renewable supply that is curtailed, whereas for all other generators (i.e., $\forall g \in \mathcal{G}_c$), they represent the amount of electricity that is over-generated in period $t$. Here, we assume the existence of a mechanism that can consume over-generation at the buses where conventional generators are located in. In more realistic settings, the over-generated electricity should be accounted for at certain locations where a consumer (e.g., a neighboring grid or energy-storage facilities) exists. Such information is not available in the NREL-118 dataset.

All conventional generators (including hydro) must obey certain physical requirements for attaining feasible production schedules. The generator capacities and minimum generation requirements are given by
\begin{equation}
\label{con:generation_capacity}
G_g^{\min} x_{gt} \leq G_{gt} \leq G_g^{\max} x_{gt}, \qquad \forall g \in \mathcal{G}_c, \, t \in [t],
\end{equation}
whereas ramping requirements are modeled as follows:
\begin{equation}
\label{con:ramping}
-\Delta G_g^{\min} \leq G_{gt} - G_{gt-1} \leq \Delta G_g^{\max}, \qquad \forall g \in \mathcal{G}_c, \, t \in [t].
\end{equation}
Above (and in the ensuing discussion) $G_{gt}$ is used to simplify exposition and defined as $G_{gt} = \usedgen_{gt} + \overgen_{gt}$. Ramping constraints \eqref{con:ramping} can be strengthened with binary variables to enhance the computational performance of MIP solvers. Our study incorporated some of the developments made in \cite{Damci-Kurt2015} and \cite{Atakan2018uc}. For the purpose of conciseness, we do not present them in here.

Electricity transmission is modeled using three sets of variables which represent the electricity flow $(F_{ij,t})$, bus voltage angles $(\theta_{jt})$, and the amount of unmet demand $(D^{\text{shed}}_{jt})$. We begin with the flow-balance equations:
\begin{align} 
\label{con:flow_balance}
\sum_{i \in \mathcal{B} : (i,j) \in \mathcal{L}} F_{ij,t} - \sum_{i \in \mathcal{B} : (j,i) \in \mathcal{L}} F_{ji,t} + \sum_{g \in \mathcal{G}_j} G^+_{gt} +  D^{\text{shed}}_{jt} = D_{jt} + R_{jt}, ~ j \in \mathcal{B}, \, t \in [t].
\end{align}
Note that the above constraint also involves reserve considerations $(R_{jt})$, which are modeled as the sum of contingency and regulation requirements. 

We consider linear (direct-current) approximations of power-flows in our models. These are given, in terms of the bus voltage-angles, as follows:
\begin{align}
& F_{ij,t} = B_{ij} (\theta_{it} - \theta_{jt}), && \forall (i,j) \in\mathcal{L}, t \in [t],  \label{con:DC_approximation} \\ 
& \theta_j^{\min} \leq \theta_{jt} \leq \theta_j^{\max}, && \forall j \in \mathcal{B}, \, t \in [t]. \label{con:bus_angle_limit}
\end{align}

Finally, transmission capacities are given by
\begin{align} 
\label{con:line_capacity}
        F^{\min}_{ij} \leq F_{ij,t} \leq F_{ij}^{\max}, \qquad \forall (i,j) \in \mathcal{L}, \, t \in [t].
\end{align}

We use $\tilde{\xi} = \{\tilde{\xi}_t\}_{t\in[t]}$ to represent stochastic process corresponding to the solar and wind output.  The realization of $\tilde{\xi}_t$ under scenario $s \in \Xi$ is given by $\xi_t^s$. The components $\xi^s_{gt}$ denote the realization of solar/wind availability under scenario $s$, for generator $g \in \mathcal{G}_r$, in period $t \in [t]$. For solar and wind generators, the forecast and actual supply time-series are assumed to be capturing the physical requirements that these generators are subject to. Accordingly, we only need to impose a solar/wind availability constraints, which are given as follows:
\begin{align}
G_{gt}^+ + G_{gt}^- = \xi_{gt}^{s} x_{gt}, \qquad \qquad \forall g \in \mathcal{G}_r, \, t \in [t].
\label{con:generation_availability}
\end{align}
This constraint implies that the amount of available renewable generation is either consumed or curtailed. Note that, since renewable generators are assumed to be committed, we have $x_{gt} = 1$.

In all models, we set $\phi_g^c = \phi_g^o = \$25$ per MW $\forall g \in \mathcal{G}$, and $\phi_j^o = \$5000$ per MW $\forall j \in \mathcal{B}$. This setup is inspired by the cost figures discussed in \cite{Penn2017} and \cite{Olson2015}, respectively. Note that conventional over-generation also involves production cost, therefore our models prioritize load-shedding, conventional over-generation, and renewable curtailment, in the given order.

\subsection{Day-ahead and Short-term Unit Commitment}
A DA-UC model is solved once every day with a $24$-hour horizon and one hour time resolution with decision epochs indexed by $[i]$.
On the other hand, the ST-UC model is solved eight times a day with a horizon of four hours and decision epochs indexed by $[j]$. In either case, the stochastic UC model is given by:
\begin{align}
\min ~ & \sum_{t \in [t^\prime]} \sum_{g \in \mathcal{G}_c^*} \big( c_g^{s} s_{gt} + c_g^{f}x_{gt} \big) + \mathbb{E} \big[ ED(\x, \tilde{\xi}) \big] \notag \\ 
\text{subject to: } ~ & \eqref{con:state_equations}-\eqref{con:minimum_down_requirements}, \nonumber \\ 
&(x_{gt}, s_{gt}, z_{gt}) \in \{0,1\}^{3}, \quad \forall g \in \mathcal{G}_c^*,  \, t \in [t^\prime]. \nonumber 
\end{align}
Here, consolidated decision vector $\x = (x_{gt}, s_{gt}, z_{gt})_{\forall g \in \mathcal{G}, t \in [t^\prime]}$ and
\begin{align}
ED(\x, \xi^s) = 
\min ~ & \sum_{t \in [t^\prime]} \Bigg( \sum_{g \in \mathcal{G}_c^*} c_g^{v} \usedgen_{gt} + \sum_{j \in \mathcal{B}} \bigg( \sum_{g \in \mathcal{G}_j\cap \mathcal{G}_c^*}  \phi^{o}_g \overgen_{gt} \nonumber \\ & \qquad \qquad  + \sum_{g \in \mathcal{G}_j\cap \mathcal{G}_r}  \phi^{c}_g \overgen_{gt} + \phi^{u}_j D^{\text{shed}}_{jt} \bigg) \Bigg) \nonumber \\ 
\text{subject to: } ~ & \eqref{con:generation_capacity}-\eqref{con:generation_availability}, \nonumber \\ 
& (\usedgen_{gt}, \overgen_{gt}) \in \mathbb{R}_+^2, \qquad \forall g \in \mathcal{G},  \, t \in [t^\prime], \nonumber \\ 
& F_{ij, t} \in \mathbb{R}, \qquad \forall (i,j) \in \mathcal{L}, \, t \in [t^\prime], \nonumber \\ 
& \theta_{jt} \in \mathbb{R},\, D^{\text{shed}}_{jt} \in \mathbb{R}_+, \qquad \forall j \in \mathcal{B}, \, t \in [t^\prime]. \nonumber
\end{align}
\sloppy
The consolidated decision vector $\y$ used in the abstract model in \S\ref{sec:decisionProcess:modelingsetup} is defined as a concatenation of real-valued decision variables in the above model. That is, $\y = ((G_{gt}^\pm,)_{\forall g \in \mathcal{G}}, (F_{ijt})_{\forall (i,j) \in \mathcal{L}}, (D_{it}^\text{shed}, R_{it}, \theta_{it})_{\forall i \in \mathcal{B}}))_{\forall t \in [t^\prime]}$. Note that the ST-UC model will additionally include constraints \eqref{eqn:stuc_fixedcommitments} and \eqref{eqn:stuc_generationDeviation} that ensure that the DA-UC commitments are honored in the ST-UC model and the difference in the generation amounts is bounded. The latter restrictions are imposed only on the generation decision variables, i.e., $(G_{gt}^\pm)_{\forall g \in \mathcal{G}, t \in [t^\prime]}$. The second-stage problem captures the dispatch corresponding to solar/wind availability under scenario $\xi^s$ that affects right-hand side of constraint \eqref{con:generation_availability}. The input $\mathbf{x}$ only affects constraints \eqref{con:generation_capacity} and \eqref{con:generation_availability}.

\subsection{Hour-ahead Economic Dispatch}
Next, we describe the stochastic ED model. The first-stage consists of all the decisions associated with the first epoch, whereas the decisions for the remaining epochs in the horizon are made in the second-stage. The resulting formulation is given below.
\begin{align}
\min ~ & \sum_{g \in \mathcal{G}} c_g^{v} \usedgen_{g1} + \sum_{j \in \mathcal{B}} \bigg( \sum_{g \in \mathcal{G}_j \cap \mathcal{G}_c}  \phi^{o}_g \overgen_{g1} + \sum_{g \in \mathcal{G}_j\cap \mathcal{G}_r} \phi^{c}_g \overgen_{g1} + \phi^{u}_j D^{\text{shed}}_{j1} \bigg) \nonumber \\ & \qquad \qquad  + \mathbb{E} \big[ ED^{\prime}((G_{g1}^\pm)_{\forall g}, \xi) \big] \notag \\ 
\text{subject to: } ~ & \eqref{con:generation_capacity} - \eqref{con:line_capacity},\eqref{eqn:ed-fixedcommitments}, \eqref{eqn:ed_generationDeviation} \text{ ($t = 1$)}, \nonumber \\
& (\usedgen_{g1}, \overgen_{g1}) \in \mathbb{R}_+^2, \qquad \forall g \in \mathcal{G},  \nonumber \\ 
& F_{ij, 1} \in \mathbb{R}, \qquad \forall (i,j) \in \mathcal{L}, \nonumber \\ 
& \theta_{j1} \in \mathbb{R},\, D^{\text{shed}}_{j1} \in \mathbb{R}_+, \qquad \forall j \in \mathcal{B}, \nonumber
\end{align} 
where, 
\begin{align}
ED^{\prime}((G_{g1})_{\forall g}, \xi^s) = \min ~ & \sum_{t \in [k] \setminus \{1\}} \Bigg( \sum_{g \in \mathcal{G}} c_g^{v} \usedgen_{gt} + \sum_{j \in \mathcal{B}} \bigg( \sum_{g \in \mathcal{G}_j\cap \mathcal{G}_c}  \phi^{o}_g \overgen_{gt} \nonumber \\ & \qquad \qquad + \sum_{g \in \mathcal{G}_j\cap \mathcal{G}_r}  \phi^{c}_g \overgen_{gt} + \phi^{u}_j D^{\text{shed}}_{jt} \bigg) \Bigg) \qquad\qquad\qquad \nonumber \\ 
\text{subject to: } ~ & \eqref{con:generation_capacity}-\eqref{con:line_capacity}, \eqref{eqn:ed-fixedcommitments}, \eqref{eqn:ed_generationDeviation} \text{ (excluding $t = 1$)}, \nonumber \\ 
& G_{gt} = \xi^s_{gt} x_{gt}, \qquad \forall g \in \mathcal{G}_r, \, t \in [k] \setminus \{1\}, \nonumber\\ 
&  (\usedgen_{gt}, \overgen_{gt}) \in \mathbb{R}_+^2, \quad \forall g \in \mathcal{G},  \, t \in [k] \setminus \{1\}, \nonumber \\ 
& F_{ij, t} \in \mathbb{R}, \qquad \forall (i,j) \in \mathcal{L}, \, t \in [k] \setminus \{1\}, \nonumber \\
& \theta_{jt} \in \mathbb{R},\, D^{\text{shed}}_{jt} \in \mathbb{R}_+, \qquad \forall j \in \mathcal{B}, \, t \in [k] \setminus \{1\}. \nonumber
\end{align}
Within the hierarchical frameworks, note that the HA-ED model also includes constraints \eqref{eqn:ed-fixedcommitments} and \eqref{eqn:ed_generationDeviation} to ensure that upstream commitment decisions are honored and the differences in the generation amounts are bounded. 

\sa{
\section{Predicting Solar and Wind Energy} \label{sec:prediction}
}
The accuracy of solar and wind forecasts depends on the quality and abundance of information collected from operation sites, and the ability of the prediction models to harness them. Modern ISOs/TSOs already receive reliable streams of data from solar and wind farms, and new mechanisms are being deployed to monitor behind-the-meter generation (e.g., by rooftop solar panels; see, for instance, \citealp{NYISO2017b}). The prediction models may directly process physical input (e.g., weather, satellite images) and convert it into power forecasts, use real-time and historical supply information to make statistical predictions, or use a combination of both \citep{Argonne2009, Tuohy2015}. The produced forecasts may have a look-ahead ranging from a few minutes to several days, depending on the nature and purpose of the forecasting model. 

The NREL-118 instance involves DA supply forecasts actual observations for each wind and solar generator. However, our framework still needs a scheme to model the distributions of these forecast supply time-series, so that scenarios can be sampled and used for stochastic programs. Additionally, we need a mechanism to dynamically update the DA forecasts (and the associated scenarios). Ultimately, as time progresses, the information gained should lead to more accurate and precise distributions of the renewable supply. This suggests, for instance, that the forecast used within the ED model for a given time period should be more reliable than that used within the DA-UC for the same time-period.

We address the first necessity with an $m$-th order vector auto-regression (VAR) model, which can be stated as
\begin{align}
\text{VAR}(m) = \bar{y}_{t + i} = \sum_{j=1}^m \Phi_j \bar{y}_{t-j+1} + \epsilon_{t+i}, \qquad \forall t, \, i = 1, 2 , \ldots \nonumber
\end{align}
where, $\bar{y}_t$ is the day-ahead supply forecast for a vector of generators, $\epsilon_{t}$ is a white-noise process with a time-invariant positive-definite covariance matrix; and $(\Phi_j)_{j=1}^m$ are autoregression coefficient matrices. In contrast to a univariate autoregressive process, a VAR model can capture both temporal and spatial correlations of the stochastic processes considered. This is especially important for modeling renewable output, as weather systems can move over large areas, simultaneously impacting multiple generators. We refer the reader to \cite{Lutkepohl2007} for further details on multivariate autoregressive processes,  \cite{Hering2015} for more advanced VAR models, and \cite{Gangammanavar2018} for a similar application of VAR models to power-system planning problems.  A general discussion incorporating predictive statistical models in SP appears in \cite{DSen2021}.

In our study, separate VAR models are used for wind and solar generators, both based on two weeks of normalized past data. For creating scenarios, we sample from the distribution of the white-noise process $\epsilon_t$, and add these sampled-deviations to the forecasts corresponding to the day ahead point forecast $(\bar{y}_t)$. This procedure is repeated from scratch for every tested day. For alternative scenario generation procedures, the reader is referred to \cite{Rios2015, Woodruff2018} for solar, and \cite{Morales2010} for wind outputs, among others. We used the \texttt{vars} and \texttt{MTS} packages in \texttt{R} \citep{vars, mts} for fitting the VAR models (with a maximum lag of 3) and sampling from the noise distributions, respectively, where further details can be found in our code repository \cite{github}.

Next, we describe our scheme to capture the evolution of uncertainty across the hierarchy and time. Note that the accuracy and precision of DA supply forecasts (and scenarios) typically deteriorate as the prediction-horizon grows. Therefore, it is crucial to evolve these to prevent unlikely inputs from becoming inputs to optimization models, thereby leading to low-quality plans. Moreover, the selected update scheme should be fast and lightweight, as it will need to be invoked, on the fly, in between optimization model runs. In this study, we consider the following update rule:
\begin{align}
\tilde{y}_{t+i} = \alpha_i \big(y_t + (\bar{y}_{t+i} - \bar{y}_{t+i-1})\big) + (1-\alpha_i) \bar{y}_{t+i}, \qquad \forall t, \, i = 1, 2 , \ldots \nonumber
\end{align}
where, $\tilde{y}_t$ and $y_t$ are the updated supply forecast and actual supply for a vector of generators, respectively, and $\{\alpha_i\}$ is a sequence of coefficients with $\alpha_i \in [0, 1], \, \forall i$, and $\alpha_i \searrow 0$. The above equation adds the forecasted increases $(\bar{y}_{t+i} - \bar{y}_{t+i-1})$ to the most-recently observed generator outputs $(y_t)$, and sets the updated forecast as a combination of these and the original DA supply forecasts $(\bar{y}_{t+i})$. The latter ensures that the updated forecasts can get close to day-ahead forecast levels, despite prediction inaccuracies around exact timings (e.g., peak wind output may be achieved at a different hour than the DA forecast).

In our study, the DA-UC uses DA supply forecasts (or scenarios, if it is a stochastic model), while ST-UC and ED use updated DA forecasts according to the above scheme, based on the time period that they are solved. We note that the first period of the ED model is based on actual observations, as it is recorded for metric calculations in the computational section. Similar to solar and wind output, the demand at each bus can be considered as a stochastic process, and the above models can be used as a basis for decision-making. However, ISOs/TSOs can often obtain very accurate load forecasts during the day, and the errors in DA load forecasts are less severe than that of the renewable resources. Due to this observation, we disregard the uncertainty associated with the loads in the ST-UC and ED problems. In the DA-UC problem, however, we consider the DA forecasts of the loads. Finally, when necessary, we use spline-interpolation to obtain time-series with finer resolution.

\Urlmuskip=0mu plus 1mu\relax
\bibliographystyle{plainnat}
\bibliography{tamingRefs}

\end{document}